\documentclass[twocolumn,epsfig,graphics,showpacs,floatfix,mathbbm]{revtex4}

\usepackage{amsmath,amsfonts,amssymb,graphics,graphicx,times,epsfig,color,bbm}
\usepackage{amsthm}
\graphicspath{{figures/}}

\newcommand{\zz}{\mathbbm{Z}}

\begin{document}

\title{Percolation, renormalization,
and quantum computing with non-deterministic gates}

\author{K.\ Kieling,  T.\ Rudolph, and J.\ Eisert}

\affiliation{
QOLS, Blackett Laboratory,
Imperial College London, Prince Consort Road, London SW7 2BW, UK\\
Institute for Mathematical Sciences, Imperial College London,
Prince's Gate, London SW7 2PE, UK
}
\date\today

\begin{abstract}
We apply a notion of static renormalization to
the preparation of entangled states for quantum computing,
exploiting ideas from percolation theory. Such a strategy yields a
novel way to cope with the randomness of non-deterministic quantum gates.
This is most relevant in the context of optical architectures,
where probabilistic gates are common, and cold atoms in
optical lattices, where hole defects occur.
We demonstrate how to efficiently construct cluster states
without the need for rerouting, thereby avoiding a massive amount of
conditional dynamics; we furthermore show that except for a single layer
of gates during the preparation,
all subsequent operations can be shifted to the final adapted
single qubit measurements. Remarkably, cluster state preparation is achieved
using essentially the same scaling in resources as
if deterministic gates were available.
\end{abstract}

\maketitle

In addition to its conceptual interest, the cluster state or one-way model of
quantum computation~\cite{Cluster}
appears to yield a highly desirable route to quantum computing
for a variety of technologies [1--7],
not least due to the clear cut distinction between
the creation and consumption of entanglement. There are generally
two approaches to preparing cluster states:
In a {\it static approach}, one can make use of a physical
setting where an underlying lattice structure is naturally given, as
in case of cold atoms in optical lattices.
Here, imperfections such as hole defects are a challenge for the
preparation of a perfect cluster. On the other hand, there is a {\it dynamic approach} of building up large-scale cluster states using
probabilistic quantum gates;
this is most promising for
architectures based on linear optical systems, optical cavities,
or optical small non-linearities [5--8].
Any such scheme requires dynamics that depend on
success or failure of the entangling gates. While
cluster state computation always requires a level of
``classical'' feed-forward -- wherein settings of
single-qubit measurement devices need to be switched
according to previously obtained outcomes -- all current
proposals for building cluster states in the dynamic approach
rely on larger (by several orders of magnitude)
amounts of the much more daunting ``active switching'' type
of feed-forward. This involves the quantum systems being
routed on demand into different possible coherent interactions with other quantum
systems, based on success or failure of previous gates. This is particularly true for the linear optical
paradigm [6--10],
which is the motivation for some of our results, although we stress they are applicable to
any such probabilistic setting.

In this work we demonstrate that the two obstacles mentioned above --
local defects and active feedforward -- can be
overcome in principle by a single strategy: The appropriate use
of a static structure together with {\it classical percolation ideas}.
In the dynamic approach, it is possible to dispense with \emph{all} of the active switching,
once small initial pieces of cluster state have been obtained. Given such small clusters, every
qubit is only involved in one probabilistic two-qubit gate, followed by one single qubit measurement. The principal idea is to use  the probabilistic gates to combine
small pieces of cluster according to a specially chosen lattice geometry.
On the percolated lattice \cite{Grimmett} a pattern of single qubit measurements
can be efficiently determined by an offline classical computation, and
universal quantum computation is attainable. Remarkably, it is possible
to achieve this complete removal of active feed-forward \emph{at essentially no cost}:
The resources required induce at most a sublinear overhead per qubit
compared to the situation of having perfect \emph{deterministic} gates at hand.
We will also present strong numerical evidence of a reduction of the
overhead to being sub-logarithmic. For
linear optical settings, we will show how the initial entangled states
can be as small as $4$-qubit cluster states, which have already been
prepared in down conversion experiments~\cite{WRR+05}.

\begin{figure}
  \includegraphics[width=7.6cm]{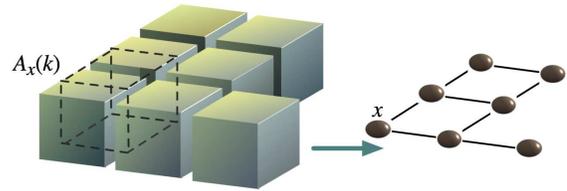}
  \caption{Renormalization procedure: Blocks $A_x(k)$ of
  the lattice $U$ (here shown with overlapping blocks using
  dashed lines) with
  crossing clusters give rise to renormalized sites
  $x\in M$.
   \label{fig:renormalization} }
\end{figure}

The technique we use to deal with the randomness of the cluster states
is that of coarse graining of an underlying lattice $U$ into blocks
corresponding to logical qubits, forming a {\it renormalized lattice}
$M$. Here, vertices comprise the blocks,
and edges reflect connections between crossing clusters in neighboring blocks,
see Fig.\ \ref{fig:renormalization}. We want $M$ to be a fully occupied lattice with
{\it asymptotic certainty}, and we seek to identify the {\it scaling} of the resources
required to achieve this.

For concreteness we focus on
$M= [1,L]^{\times 2}$ for some length $L$, that is, the renormalized
lattice is a 2d cubic (square) lattice. Of this, a hexagonal sublattice
will be used for quantum computing, the graph states of
which constitute universal resources. We focus on bond percolation,
so a bond is present (``open'') with probability $p$.
The underlying lattice has to be chosen such that
in some way it is possible to exceed the
critical probabilities, marking the arrival of infinite connections throughout the lattice \cite{Grimmett},
with the initial resources and gates at hand.
When $p$ already exceeds the critical bond-percolation probability $p_{\text c}$ of a
{\it two-dimensional lattice} (e.g., $1/2$ for the square lattice),
this natural geometry can be introduced.
Then, a possible renormalization amounts to simply exploiting
vertical and horizontal paths which by standard results necessarily have to
cross sufficiently many times.
Our goal is to go significantly
further and deal also with the situation of small $p$:
Techniques to increase $p$ (e.g., replace each
bond by multiple ones in parallel to increase the
probability that at least one of them exisits) or
to decrease $p_{\text c}$ (using another lattice with
higher threshold, notably ones in {\it three or higher dimensions}
to generate a 2d lattice) will be used.
This will in general increase the vertex degree, causing a trade-off between the number and the size of the initial cluster state resources.
Within all solutions which provide efficient scaling of the resource number, the main interest will be to minimize
their sizes.

In the most prominent context at hand, namely
fusion gates \cite{BR04} operating with a success
probability of at most $1/2$, we will overcome this problem
by taking $U\subset \zz^3$,
so starting from a 3d cubic lattice, for which
$p_{\text c}= 0.249$. We will identify each vertex  $x\in M$
with a block of size $(2k)^{\times 3}$.
We can meaningfully define an {\it event}
$\mathfrak{A}_x(k)$ of $x\in M$ being
``occupied''. With this we mean that
there exists a crossing open cluster within the block,
so a connected path on the graph
connecting each pair of faces on opposite sides,
at least in the first and second dimension \cite{Grimmett}.
Moreover, this crossing cluster is connected to each of the ones
of the blocks associated with sites $y$ adjacent to $x$, which does not arise
as naturally as in the 2d case.
We show the following:

\begin{figure}
  \includegraphics[width=160pt]{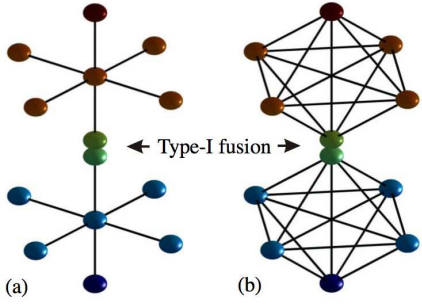}
  \caption{(a) Placing 7-qubit clusters at the vertices of a cubic lattice and implementing
    a probabilistic parity check gate (such as a linear optical
    Type-I fusion gate \cite{BR04})
    results in a percolated cluster. (b) For
    quantum computation it suffices to use
    the $6$-qubit
    graph states \cite{Cluster,Graphs} depicted (i.e., the
    complete graph $K_6$)
    forming the covering lattice.
 \label{fig:7stars} }
\end{figure}

 {\it Renormalized cubic lattices:}
 {\it  Let $p>p_{\text c}$. Then
for any $\mu>0$,
the probability $P_p(L)$ of having $\mathfrak{A}_x(k)$ satisfied
for all $x\in M$ with $k=L^\mu $ fulfills}
\begin{equation}\label{eq1}
    \lim_{L\rightarrow\infty} P_p(L) = 1.
\end{equation}
In other words, with a sublinear overhead $k=O(L^\mu)$,
one can create a cubic lattice $M=[1,L]^{\times 2}$
out of $U$ using {\it bond percolation}.
Moreover, this preparation is
asymptotically {\it certain} (in the same sense as in Refs.\ \cite{KGE06}),
despite the underlying elements
being probabilistic.
The value of $k$ specifies to what extent we ``dilute'' the
superlattice $M$ compared to $U$.

To show the validity of
(\ref{eq1}), we introduce a series of
blocks of the underlying lattice $U$,
which, in addition to the blocks of $M$ include blocks overlapping with those (see dashed lines in Fig.~\ref{fig:renormalization}).
For any $y\in [2 ,2L]^{\times 2}$,
let
$A_{y}(k) =
    [  y_1 k, y_1 k + 2k-1 ]
    \times
    [  y_2 k, y_2 k + 2k -1]
    \times  [1,2k]$ \cite{Grimmett}.
    Each vertex $x\in M$ is identified
    with $y=2x$. To show that  $P_p(\mathfrak{A}_x(k))=1$ (almost certainly)
    for all $x\in M$ for large $L$,
    we make use of statements on crossing
clusters in cubic lattices, as well as of a convenient
tool in percolation theory,
the {\it FKG inequality}: Let $\mathfrak{B}$ and $\mathfrak{C}$ be two {\it
increasing events}, i.e., events that ``become more likely'' for
increasing $p$. Then the FKG inequality states that
$P_p(\mathfrak{B} \cap \mathfrak{C}) \geq P_p(\mathfrak{B}) P_p(\mathfrak{C})$ \cite{Grimmett}. In other
words, increasing events are positively correlated.

Let us denote with $\mathfrak{B}_y(k)$ the event that $A_y(k)$ has a
left-to-right crossing cluster in the first dimension, i.e.,
an open path having vertices $a$ and $b$ satisfying
$a_1 =  y_1 k$ and $b_1 =  y_1 k + 2 k -1$.
Now there exists a constant
$d>0$, only dependent on $p$, such that
$ P_p(\mathfrak{B}_y(k))  \geq 1 - \exp(-d k^2)$
for $k\geq 3$ \cite{Grimmett}. We only need to ``connect these vertices''.
The blocks $A_y(k)$ and $A_z(k)$
are overlapping for $\text{dist}(y,z)=1$.
Now take two sites $y\in[2,2L-1]\times[2,2L]$, with $y_2$ even,
and $z$ with $z_1=y_1+1$, and $z_2=y_2$.
The events $\mathfrak{B}_y(k)$ and $\mathfrak{B}_z(k)$
are increasing events, and
therefore, we can use the FGK inequality: intuitively,
if in $A_y(k)$ there is already a crossing cluster, then this
crossing cluster is already half way through $A_z(k)$, and
hence renders a crossing cluster there more likely.
Consider the overlap
between two adjacent blocks, $B_y(k)= A_y(k)\cap A_z(k)$.
We can define the following event: For $p\in[0,p_{\text c}]$,
we define $\mathfrak{D}_y(k)$ as the event
that never occurs, for $p\in (p_{\text c},1]$ it is the event of
having {\it at most a single} left-to-right
crossing cluster in $B_y(k)$.
This is an increasing event \cite{Aizenman97}.
Hence, the probability of having simultaneously a left-to-right crossing cluster in
$A_y(k)$, one in $A_z(k)$, and exactly one in $B_y(k)$ can be
estimated using the FKG inequality.
There exist constants
$c,a>0$, only dependent on $p$ such that the
probability of having the event $\mathfrak{D}_y(k)$ satisfies \cite{Aizenman97}
\begin{equation}
    P_p(\mathfrak{D}_y(k)) \geq 1- (2k)^6 a \exp(- c k).
\end{equation}
So, using again the FKG inequality,
one
finds that the probability of $\mathfrak{E}_{y}(k)$
of having two crossing clusters
in $A_y(k) $ and $A_{(y_1+1,y_2)}(k)$ which are actually connected as
$ P_p(\mathfrak{E}_y(k)) \geq (1 - \exp(-d k^2))^2 (1 - (2k)^6 a\exp(- c k))$.
This procedure can be iterated, using FKG in each step.
To find connections in the other direction, we re-use
the argument on having at most a single crossing cluster,
but now using $[1,4k]$ in the third direction
to be able to apply the results of Ref.~\cite{Aizenman97}.
This gives an overall probability of having $\mathfrak{A}_x(k)$ for each
$x\in M$ of
\begin{eqnarray}
  &P_p(L,k)=P_p(\cup_x\mathfrak{A}_x(k)) \geq  (1 - \exp(-d k^2))^{2L^2-L}\times& \nonumber \\
   &\left(
   (1-(2k)^6 a\exp(- c k))^2
   (1-(4k)^6 a\exp(- c 2k))\right)^{L(L-1)}.& \nonumber
\end{eqnarray}
We bound this expression from below with the slowest increasing term, i.e.,
there exists an integer $k_0$ such that
$P_p(L,k) \geq (1-(2k)^6 a\exp(- c k))^{5L^2}$
for all $k\geq k_0$. Let us set $k=  L^\varepsilon$ for $\varepsilon>0$.
Then, $\lim_{L\rightarrow\infty} P_p(L,k(L))=1$ using that for any $e,f>0$, we have that
$\lim_{n\rightarrow\infty}\left( 1 - e n^{3 \varepsilon} \exp(-f n^{\varepsilon/2}) \right)^{n}=1$. This means
that by using a sublinear overhead, we arrive at
an asymptotically \emph{certain} preparation of the renormalized
lattice.

This gives rise to an overall resource requirement of
$O(L^\varepsilon)^3\times L^2=O(L^{2+3\varepsilon})$ $7$-qubit
states to build a fully connected cluster state that
(almost certainly) consists of $L\times L$
blocks, and requires no rerouting.
As long as $p>p_{\text c}$, this scaling will hold.
This should be compared to the $O(L^2)$ qubits we
would require if we had perfect deterministic gates with which to
build the cluster.

To utilize the renormalized blocks some classical computation is needed,
and we need to ascertain that it is efficient in the system size.
One first has to find the crossing clusters in each block, e.g., by the
\emph{Hoshen-Kopelman-algorithm}~\cite{HK76} ($O(k^3)$ steps, $O(k^2)$ memory).
Then, using a series ($O(L^2)$) of breadth-first-searches ($O(k^3)$) equips us with
loop-free connections between chosen ``mid-points'' of all sets of neighboring blocks.
For conceptual simplicity it suffices to only identify $3$-way ``T-junctions'' (instead of crosses),
and use these to build a hexagonal lattice, which can be converted to a square lattice easily \cite{Nest}.
The unused qubits in the blocks can be eliminated using $\sigma_z$ measurements,
all unwanted qubits on the connections between the T-junctions can be cut out
with $\sigma_x$ and $\sigma_y$ measurements. Thus, the amount of classical computation
needed shows the same scaling behavior as the number of required resources \cite{QIIC}.

At this stage we have used $7$-qubit clusters on a cubic lattice,
see Fig.\ \ref{fig:7stars}.
We now turn to various methods for reducing the size of this initial resource.
The first one is quite general, and will apply to any lattice. We see from
Fig.~\ref{fig:7stars} that a qubit is left on each successfully formed edge.
One interesting observation is that this qubit may be measured out,
relaxing the requirement of photon number resolving detectors to dichotomic
detectors. However, one might also use this to construct the
\emph{covering lattice} \cite{Grimmett}
of the original lattice, by connecting these sites with all perimeter sites
from the neighboring stars, and removing the stars' central qubits
(Fig.~\ref{fig:7stars}(b)). From percolation theory it is
known~\cite{Grimmett} that the critical
bond percolation probability of a lattice equals the critical
site percolation probability of the covering lattice (for which a site
is ``open'' with a certain probability $p$).
Thus by using $6$-qubit clusters (with the
connectivity of the complete graph $K_6$ as shown)
the covering lattice can be built by fusion of neighboring corner qubits. These percolation processes are equivalent for our
purposes, because a path between two arms of one star in the original lattice exists iff the fusion processes involving these two arms were successful, and a path between two corner qubits in the covering lattice exists iff the
fusion attempts on the equivalent two qubits were successful.

%
\begin{figure}
  \includegraphics[width=140pt]{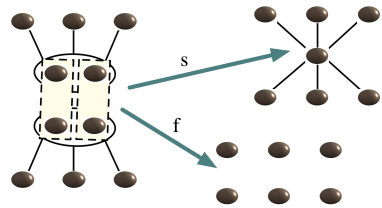}
  \caption{A pair of $5$-qubit states (star with $3$ arms, central qubit redundantly encoded) can be used to create a single $7$-qubit GHZ state with $p=3/4$.
      A Type-I and a Type-II fusion are applied to the centers.
      On success of one fusion gate, the centers
      merged into a single redundantly
      encoded qubit and subsequent application of another
      fusion gate will succeed and only reduce
      the level of encoding. If the first fusion fails the second one may still succeed with $p=1/2$. As the order does not matter, both gates may be applied
      simultaneously, without any need for rerouting.
      The bond fusion processes are not influenced, only the site probability is determined by this process.
    \label{fig:5cluster_basic}}
\end{figure}

A quite different method (somewhat more specific to linear optics)
can further reduce the size of the initial states on the cubic lattice to stars with $3$ arms where the central qubit is redundantly encoded by
judiciously fusing the two ``central'' qubits of each of the $5$-stars, while simultaneously applying the Type-I fusion operations on the bonding qubits (see Fig.\ \ref{fig:5cluster_basic}).
A Type-I and a Type-II fusion are applied to the centers.
On success of one fusion gate, the centers merged into a single redundantly
encoded qubit and subsequent application of another fusion gate will succeed and only reduce the level of encoding.
If the first fusion fails the second one may still succeed with $p=1/2$. As the order does not matter, both gates may be applied
simultaneously, without any need for rerouting, so site-preparation succeeds with probability $3/4$.
If the central (``site'') fusion fails, the bond fusions can still be attempted as usual.
The single qubits resulting from the failure are in the state $|+\rangle^{\otimes6}$, and
fusion gates involving them will succeed or fail with probability $1/2$.
Hence, the site and edge generation processes are independent and do not require active switching,
allowing for application of the \emph{mixed} percolation model~\cite{Grimmett}.

A more general approach to decrease the size of the initial resources is the following \cite{Decrease}: 
Instead of using the cubic lattice, we switch to a different 3d lattice, e.g., the one 
with the lowest vertex degree, namely
the diamond lattice which has vertex degree 4, and a bond
percolation threshold of $p_{\text c}=0.389$.
While percolating on the diamond lattice directly would require 5-qubit star clusters, 
by percolating on the covering lattice (as explained above),
the \emph{pyrochlore lattice},
we even further reduce the initial resources required to $4$-qubit GHZ states,
which lies below the resource size dictated by a naive ansatz with a square lattice!
These tetrahedra consist of triangles and are thus not two-colorable.
However, it can readily be shown that
the resulting graph states can still be reduced
to universal cluster states.

\begin{figure}
  \includegraphics[width=8cm]{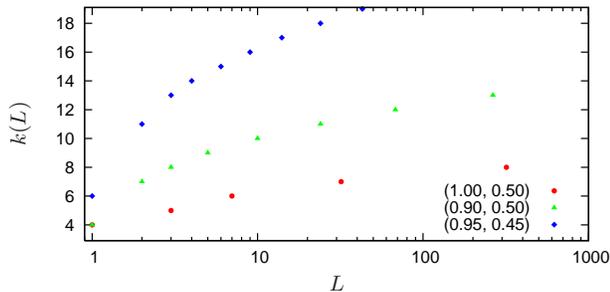}
  \caption{Dependence of the diamond lattice block size $k^{\times 3}$ on the size $L$ of the renormalized square lattice for
    for three different sets of site- and bond probabilities $(p_{\text{site}},p_{\text{bond}})$.
    The overall success probability threshold $P(L)$ was chosen to be $1/2$.
    $10^5$ blocks of each size were
    created; each lattice site was randomly populated $10^3$ times.
    \label{fig:diamond_thresholds} }
\end{figure}

As less is known analytically about percolation for
the diamond lattice, we have turned to a numerical
verification that this lattice suffices for our purposes.
In fact, we find that the resource scaling appears slightly
more favorable than the upper bound proven
above for the cubic lattice.
Cubic blocks of the diamond lattice of size $k^{\times 3}$ have
been simulated
and
arranged in two dimensions as described above, then used
as renormalized lattice.
These sites are occupied iff there exist crossing clusters connecting the four faces.
Bonds between neighboring sites exist iff the crossing clusters of the corresponding
blocks
are connected through the common face. Depending on $k$ and the
probabilities of a site and an edge being open,
the probability $P(L)$ of building up the
whole renormalized lattice of size $L\times L$ without any
missing sites or bonds is obtained. By requiring a
fixed threshold $P$, the scaling of the block size $k(L)$
that is needed to lie above this threshold is found. The results are summarized in
Fig.~\ref{fig:diamond_thresholds}, which suggests a scaling of
$k(L)=o(\log(L))$ for each set of parameters,
thus a scaling of $L^2o(\log^3(L))$ of $4$-qubit cluster states
to build a lattice of size $L\times L$ with a success probability of at least $P$.

Although the primary concern of this work is to show how to effectively
remove active feed-forward and to deal with probabilistic gates or
configurations using percolation tools, we briefly address the role of
further imperfections in such a setting. To start with, 
losses, the occurrence of which is known without the need for destructively measuring them
(e.g., atomic systems with probabilistic entangling gates \cite{Other}),
can be accounted for by measuring out the qubits around failure sites in
the $\sigma_z$-basis. This is effectively a new percolating model, albeit
with correlated probabilities. Numerical investigations show that a similar reasoning
as above is expected to hold, demonstrating
that losses of $10\%$ can easily be accounted for.
More importantly, concerning losses like photon loss, which are 
only detected by destructive qubit measurements,
this scheme is no different than others, i.e., standard techniques
can make it loss tolerant, although, needless to say, with significantly more effort.
Our approach readily suggests two strategies to
cope with such errors:  On the one hand, by 
fixing the block size, one would fix the effective (non-heralded)
site occupation probability. 
Then, schemes for fully-fletched fault-tolerant one-way
computation can be used, once above the respective fault-tolerant 
threshold \cite{FT}. On the other hand, to suppress loss rates,
specifically photon loss, it is legitimate to consider initial 
encodings like tree-structures \cite{FT}. They in turn can be grown
probabilistically as well, still without any need for active switching.
For example, a crude estimate is the following: To correct for these losses 
occuring with a rate of $10\%$, trees with branching parameters
$(6,7,7,1)$ can be used to suppress the loss to an effective rate of $10^{-5}$. 
Together with blocks of size $6^{\times3}$
elementary diamond cells (which fail to be 
crossed with a probability of $4.76\cdot10^{-6}$),
the overall site loss rate on the renormalized lattice lies 
below the $3\cdot10^{-3}$\cite{FT} limit \cite{Depth}.

In this work we have introduced a method based on percolation phenomena of building
cluster states with probabilistic entangling gates.
The scheme dramatically reduces the amount of coherent feed-forward required; specifically there are no rerouting steps needed, once one
starts from appropriate building blocks which can be as small as $4$-qubit states.
We provided a proof that
to prepare an $L\times L$ cluster state, asymptotically with certainty,
even with this very restricted set of tools,
a scaling in the number of resources of $O(L^{2+\varepsilon})$
for any $\varepsilon>0$ can be achieved.
Numerical simulations have been carried out,
suggesting an even better resource consumption
of $L^2 o(\log^3(L))$, which should be compared to $L^2$
in the case of perfect deterministic gates.

We emphasize that while we have focussed on a rigorous proof of the utility of one type of percolation model,
there are many obvious ways in which our work can be extended: different lattices, site percolation,
dynamically growing percolated clusters and finding optimal, 
loss corrected paths ``on the fly'', and much more. We expect therefore our ideas to be applicable to a wide range of architectures
where probabilistic quantum gates originate, e.g.,
from exploiting {\it small non-linearities} as in Ref.\ 
\cite{Munro}, or linear optics \cite{BR04}.
In the context of ultracold atoms in optical lattices,
cluster states may be prepared by exploiting
{\it cold collisions} \cite{MGW+03}. One could then think of universal
computational resources when starting with a Mott state
exhibiting {\it hole defects}, such that the filling is not exactly
a single atom per site, giving rise to a site percolation variant of our argument.
It would also be interesting to see
whether the new freedom of measurement-based
schemes for quantum computing beyond the one-way
computer \cite{GE} gives rise to
further improvements concerning resource requirements.
The presented ideas
open up a new way to deal with randomness of
probabilistic gates in quantum computing.

We thank G.~Pruessner, T.J.\ Osborne, and M.~Varnava, and
W.\ D{\"u}r for discussions and
the DFG (SPP 1116), the EU (QAP), the
EPSRC, the QIP-IRC, Microsoft Research through the European PhD Programme,
EURYI, and the US Army Research Office (W911NF-05-0397) for support.


\begin{thebibliography}{99}

\bibitem{Cluster}
    R.\ Raussendorf and H.-J.\ Briegel,
    Phys.\ Rev.\ Lett.\ {\bf 86}, 5188 (2001);
    R.\ Raussendorf, D.E.\ Browne, and H.J.\ Briegel,
    Phys.\ Rev.\ A {\bf 68}, 022312 (2003).

\bibitem{Graphs}
    M.\ Hein et al., quant-ph/0602096;
   M.\ Hein, J.\ Eisert, and H.J.\ Briegel, ibid.\
   {\bf 69}, 062311 (2004).

\bibitem{MGW+03}
    O.\ Mandel et al., Nature {\bf 425}, 937 (2003).

\bibitem{Other}
    J.\ Cho and H.-W.\ Lee,
    Phys.\ Rev.\ Lett.\ {\bf 95}, 160501 (2005);
    P.\ Dong et al., quant-ph/0511045.
    Y.L.\ Lim, S.D.\ Barrett, A.\ Beige, P.\ Kok, and L.C.\ Kwek,
    Phys.\ Rev.\ A {\bf 73}, 012304 (2006).

\bibitem{Munro}
    S.G.R.\ Louis et al.,
    New J.\ Phys.\ {\bf 9}, 193
    (2007).

\bibitem{Nielsen04}
    M.A.\ Nielsen, Phys.\ Rev.\ Lett.\ {\bf 93}, 040503 (2004).

\bibitem{BR04}
    D.E.\ Browne and T.\ Rudolph,
    Phys.\ Rev.\ Lett.\ {\bf 95}, 010501 (2005).

\bibitem{KGE06}
    K.\ Kieling, D.\ Gross, and J.\ Eisert,
    J.\ Opt.\ Soc.\ Am.\ B {\bf 24}(2), 184 (2007);
    D.\ Gross, K.\ Kieling, and J.\ Eisert,
    Phys.\ Rev.\ A {\bf 74}, 042343 (2006).

\bibitem{FransonFF}
    T.B.\ Pittman, B.C.\ Jacobs, and
    J.D.\ Franson, Phys.\ Rev.\ A {\bf 66}, 052305 (2002).

\bibitem{KLM}
    E.\ Knill, R.\ Laflamme, and G.J.\
    Milburn, Nature {\bf 409}, 46 (2001);
    S.\ Scheel and N.\ L{\"u}tkenhaus,
    New J.\ Phys.\ {\bf 6}, 51 (2004); J.\ Eisert,
    Phys.\ Rev.\ Lett.\ {\bf  95}, 040502 (2005).

\bibitem{Grimmett}
    G.\ Grimmett, {\it Percolation} (Springer, Berlin, 1999);
    D.\ Stauffer and A.\ Aharony, {\it Introduction to percolation theory}
    (Taylor and Francis, London, 1994).

\bibitem{WRR+05}
    P.\ Walther et al., Nature {\bf 434}, 169 (2005).

\bibitem{Nest}
    M.\ Van den Nest et al.,
    Phys.\ Rev.\ Lett.\ {\bf 97}, 150504 (2006).

\bibitem{QIIC}
    An implementation of this procedure is provided on the website
    http://www3.imperial.ac.uk/quantuminformation.

\bibitem{Aizenman97}
    M.\ Aizenman, Nucl.\ Phys.\ B {\bf 485}, 551 (1997).

\bibitem{HK76}
    J.\ Hoshen and R.\ Kopelman,
    Phys.\ Rev.\ B {\bf 14}, 3438 (1976).

\bibitem{Decrease}
	It appears that decreasing the size of the initial states to 
	using \emph{entangled pair resources} (EPR) 
	as bonds and ``gluing'' them at sites does not 
	work, as the probability of such an operation to 
	succeed seems to drop exponentially
	with the number of qubits.

\bibitem{FT}
    C.M.\ Dawson, H.L.\ Haselgrove, and M.A.\ Nielsen,
    Phys.\ Rev.\ Lett.\ {\bf  96}, 020501 (2006);
    M.\ Varnava, D.E.\ Browne, and T.\ Rudolph,
    ibid.\ {\bf 97}, 120501 (2006); R.\ Raussendorf and J.\ Harrington,
    ibid.\ {\bf 98}, 190504 (2007);  
    P.P.\ Rohde, T.C.\ Ralph, and W.J.\ Munro, 
    quant-ph/0701090.

\bibitem{Depth}	
	In-depth analysis of 
	techniques to cope with loss 
	 will be the subject of further investigation.
 
\bibitem{GE}
    D.\ Gross and J.\ Eisert, Phys.\ Rev.\ Lett.\ {\bf 98},
    220503 (2007).

\end{thebibliography}
\end{document}